\documentclass[twocolumn,showpacs,preprintnumbers,amsmath,amssymb,nofootinbib]{revtex4}
\input psfig.sty
\usepackage{epsf}
\usepackage{graphicx,color}  
\usepackage{dcolumn}   
\usepackage{bm}        

\newcommand{\be}{\begin{equation}}
\newcommand{\en}{\end{equation}}
\newcommand{\bea}{\begin{eqnarray}}
\newcommand{\ena}{\end{eqnarray}}

\begin{document}

\title{Cosmographic constraints on a class of Palatini $\mathbf{f(R)}$ gravity}

\author{N. Pires$^1${\footnote{npires@dfte.ufrn.br}}}

\author{J. Santos$^1${\footnote{janilo@dfte.ufrn.br}}}

\author{J. S. Alcaniz${^2}${\footnote{alcaniz@on.br}}}

\affiliation{$^1$Departamento de F\'{\i}sica, Universidade Federal do Rio G. do Norte,  59072-970 Natal - RN, Brasil}

\affiliation{$^2$Observat\'orio Nacional, 20921-400 Rio de Janeiro - RJ, Brasil}

\date{\today}


\begin{abstract}

Modified gravity, known as $f(R)$ gravity, has presently been applied to Cosmology as a realistic alternative to dark energy. For this kind of gravity the expansion of the Universe may accelerate while containing only baryonic and cold dark matter. The aim of the present investigation is to place cosmographic constraints on the class of theories of the form $f(R)=R - \alpha/R^n$ within the Palatini approach. Although extensively discussed in recent literature and confronted with several observational data sets, cosmological tests are indeed inconclusive about the true signal of $n$ in this class of theories. This is particularly important to define which kind of corrections (infra-red or high-energy) to general relativity this class of theory indeed represent. We shed some light on this question by examining the evolution of the deceleration parameter $q(z)$ for these theories. We find that for a large range of $\alpha$, models based on $f(R) = R - \alpha/R^{n}$ gravity in th
 e Palatini approach can only have positive values for $n$, placing thus a broad restriction on this class of gravity.

\end{abstract}


\maketitle


\section{Introduction}


$f(R)$ gravity examines the possibility of modifying Einstein's general relativity (GR) by adding terms proportional to powers of the Ricci scalar $R$ to the Einstein-Hilbert Lagrangian \cite{Starobinsky}. The cosmological interest in these theories comes from the fact that they can  naturally exhibit an accelerating expansion phase of the universe without introducing dark energy. However, the freedom in the choice of different functional forms of $f(R)$ gives rise to the problem of how to constrain the many possible $f(R)$ gravity theories on theoretical and/or observational grounds. In this regard, much efforts have been developed so far, mainly from the theoretical viewpoint~\cite{Ferraris} (see also Refs.~\cite{Francaviglia} for recent reviews). General principles such as the so-called energy conditions~\cite{energy_conditions}, nonlocal causal structure~\cite{RSantos}, have also been taken into account in order to clarify its subtleties. More recently, observational cons
 traints from several cosmological data sets have been explored for testing the viability of these theories~\cite{Amarzguioui,Tavakol,Janilo,Fabiocc,Yang,Li,Koivisto,Fairbairn}.

An important aspect that is worth emphasizing concerns the two different variational approaches that may be followed when one works with $f(R)$-gravity theories, namely, the metric  and the Palatini formalisms (see, e.g., \cite{Francaviglia}). In the metric formalism the connections are assumed to be the Christoffel symbols and the variation of the action is taken with respect to the metric, whereas in the Palatini variational approach the metric and the affine connections are treated as independent fields and the variation is taken with respect to both. In fact, these approaches are equivalents only in the context of GR, i.e., in the case of linear Hilbert action; for a general $f(R)$ term in the action, they provide completely different theories, with almost no similarities and very distinct equations of motion\footnote{These differences also extend to the observational aspects of these approaches. For instance, we note that cosmological models based on a power-law function
 all form in the metric approach fail in reproducing the standard matter-dominated era followed by an acceleration phase~\cite{Amendola}, whereas in the Palatini approach, analysis of a dynamical autonomous systems for the same Lagrangian density have shown that such theories admit the three post inflationary phases of the standard cosmological model~\cite{Tavakol}.}. 

In this \emph{brief report} we will restrict ourselves to the Palatini formalism. By discussing the deceleration/acceleration history of the Universe through the evolution of the deceleration parameter $q(z)$, we place stringent bounds on a class of $f(R)$ theory given by $f(R) = f(R) = R - \alpha/R^n$. In particular, we find that the range of negative values of $n$ predicts a behavior that is grossly inconsistent with cosmological observations (e.g. CMB~\cite{WMAP-7}), being therefore ruled out even if it is compatible with supernova, lensing statistics and local gravity constraints.


\section{Palatini $\mathbf{f(R)}$ cosmologies}  \label{Palatini-app}


$f(R)$ cosmologies are based in the modified Einstein equations of motion derived from the dubbed $f(R)$ gravity.
The action that defines an $f(R)$ gravity is given by
\begin{equation}
\label{actionJF}
S = \frac{1}{2\kappa^2}\int d^4x\sqrt{-g}f(R) + S_m\,,
\end{equation}
where $\kappa^2=8\pi G$, $g$ is the determinant of the metric tensor and $S_m$ is the standard action for the matter
fields. Treating the metric and the connection as completely independent fields, variation of this action with respect
to the metric provides the field equations
\begin{equation}
\label{field_eq}
f'R_{(\mu\nu)} - \frac{f}{2}g_{\mu\nu}  = \kappa^2T_{\mu\nu}\,;
\end{equation}
while variation with respect to the connection gives
\begin{equation}  \label{connections_eq}
\tilde{\nabla}_\beta\left( f'\sqrt{-g}\,g^{\mu\nu}\right)=0\,.
\end{equation}
In (\ref{field_eq}) $T_{\mu\nu}$ is the matter energy-momentum tensor which, for a perfect-fluid, is given by
$T_{\mu\nu} = (\rho_m + p_m)u_{\mu}u_{\nu} + p_m g_{\mu\nu}$,
where $\rho_m$ is the energy density, $p_m$ is the fluid pressure and $u_{\mu}$
is the fluid four-velocity. Here, we adopt the notation $f'=df/dR$ and
$f''=d^2f/dR^2$. The second equation (\ref{connections_eq}) give us the connections $\Gamma_{\mu\nu}^{\rho}$,
which are related with the Christoffel symbols $\left\{^{\rho}_{\mu\nu}\right\}$ of the metric $g_{\mu\nu}$ by
\begin{equation}  \label{connections}
\Gamma_{\mu\nu}^{\rho} = \left\{^{\rho}_{\mu\nu}\right\} + \frac{1}{2f'}\left( \delta^{\rho}_{\mu}\partial_{\nu} +
\delta^{\rho}_{\nu}\partial_{\mu} - g_{\mu\nu}g^{\rho\sigma}\partial_{\sigma} \right)f'\,.
\end{equation}
Note that in (\ref{field_eq}), $R_{\mu\nu}$ must be calculated in the usual way, i.e., in terms of the independent connection
$\Gamma_{\mu\nu}^{\rho}$, given by (\ref{connections}) and its derivatives.

We assume a homogeneous and isotropic Friedmann-Lema\^{i}tre-Robertson-Walker  universe whose metric is
$g_{\mu\nu}=diag(-1,a^2,a^2,a^2)$, where $a(t)$ is the cosmological scale factor.
The generalized Friedmann equation, obtained from (\ref{field_eq}), can be written in terms of the redshift parameter $z=a_0/a -1$
and the density parameter $\Omega_{mo} \equiv \kappa\rho_{mo}/(3H_0^2)$ as 
\begin{equation}
\label{fe3}
\frac{H^2}{H_0^2} = \frac{3\Omega_{mo}(1 + z)^3 + f/H_0^2}{6f'\left(
1 + \frac{9}{2}\,\frac{f''}{f'}\,\frac{H_0^2\Omega_{mo}(1+z)^3}{Rf'' - f'}\right)^2}\,,
\end{equation}
where $\rho_{mo}$ is the matter density today. In terms of these quantities, the trace of Eq. (\ref{field_eq}) provides an important relation:
\begin{equation}
\label{trace2}
Rf' - 2f = -3H_0^2\Omega_{mo}(1 + z)^3\,.
\end{equation}


\section{Cosmographic constraints}  \label{kinematical-constr}


Cosmographic parameters linked to geometry play a key role in cosmology. In particular, the deceleration parameter, defined as $q(a) = -a\ddot{a}/\dot{a}^2$, can be written in terms of $H(z)$ and its derivative as
\begin{equation}  \label{q(z)}
q(z)={1\over H(z)}{dH \over dz}(1+z) -1\,.
\end{equation}
In what follows we use (\ref{fe3}) and (\ref{q(z)}) to place constraints on the free parameters ($\alpha,n$) of one of the most extensively discussed $f(R)$ theory, i.e.,
\begin{equation}  \label{first-f}
f(R) = R - \frac{\alpha}{R^n}\,.
\end{equation}
Note that positive values of $n$ provide IR modifications of GR while negative ones correspond to high-energy corrections. For $n=0$ this kind of gravity reduces to Einstein's general relativity plus a cosmological constant, $f(R) = R - 2\Lambda$. It is worth noting that, for a theory like (\ref{first-f}), Eq. (\ref{trace2})  evaluated at $z=0$ imposes the following relation between $\Omega_{mo}$ and the pair ($n,\alpha$)
\begin{equation} \label{alpha-relation}
\alpha = \frac{R_0^{n+1}}{n+2}\,\left( 1 - \frac{3\Omega_{mo}H^2_0}{R_0} \right)\,,
\end{equation}
where $R_0$ is the value of the Ricci scalar today. Note also that $R_0$ can be determined from the algebraic equation resulting from equating (\ref{fe3}) and (\ref{trace2})  for $z=0$. Hence, specifying the values of two of these parameters the third is automatically fixed. In other words, in the Palatini approach, one of the two free parameters $n$ and $\alpha$ can be eliminated by the constraint (\ref{alpha-relation}).

\begin{table}[t]   \label{n-Table}
\begin{center}
\begin{tabular}{lccc}
\hline \hline \\
Test& Ref. & $n$ & $\alpha$\\
\hline \hline \\
SNe Ia ({\emph{Gold}})\footnote{+ CMB + BAO}  & \cite{Amarzguioui} & [-0.3,0.3] & [-6.8,-2.2]\\
SNe Ia (SNLS)$^a$ & \cite{Tavakol} & [-0.23,0.42] & [2.73,10.6]\\
SNe Ia (\emph{Union})$^a$ & \cite{Janilo} & [-0.3,0.1] & [1.3,5.5] \\
$H(z)$$^a$ & \cite{Fabiocc} & [-0.25,0.35] & [2.3,7.1]\\
Strong lensing & \cite{Yang} & [-0.202,0.078] & [2.89,4.67]\\
SNLS + WMAP + SDSS\footnote{ + BAO + matter power spectrum + TT CMB spectrum} & \cite{Li} & \quad [-5.98,-2.12]$\times 10^{-6}$ &  \\
SDSS\footnote{+ CMB + matter power spectrum}  & \cite{Koivisto} & \quad [-2.321,1.329]$\times10^{-5}$ &  \\
\hline \hline
\end{tabular}
\end{center}
\vspace{-.2truecm}
\caption{Current constraints on $n$ and $\alpha$ for a $f(R)$ gravity given by (\ref{first-f}). The $\Lambda$CDM model corresponds to $n = 0$ and $\alpha=4.38$).}
\end{table}

Recently, the observational viability of the functional form (\ref{first-f}) have been discussed by several groups. Constraints from distance measurements to type-Ia Supernova (SNe Ia)~\cite{Amarzguioui,Tavakol,Janilo}, $H(z)$ determinations by differential age method~\cite{Fabiocc}, strong lensing statistics~\cite{Yang}, matter power spectra~\cite{Li}, Baryon Acoustic Oscillation peak (BAO) and CMB shift parameter have been derived providing the following intervals $n\in[-0.3,0.42]$ and $\alpha\in[1.3,10.6]$, at $99.7\,\%$ (C.L.) (see Table 1).

However, a glance through Table 1 shows that cosmological tests are indeed inconclusive about the true signal of $n$.  This in turn gives rise to the following question: what kind of corrections to GR do theories like (\ref{first-f}) indeed represent? The answer to this question surely depends of the signal of the parameter $n$. In what follows, we show that some light can be shed on this discussion just by examining the signal of $q(z)$, as given by (\ref{q(z)}) and taking into account (\ref{fe3})--(\ref{trace2}).

\begin{figure}[t]
\centerline{
{\psfig{figure=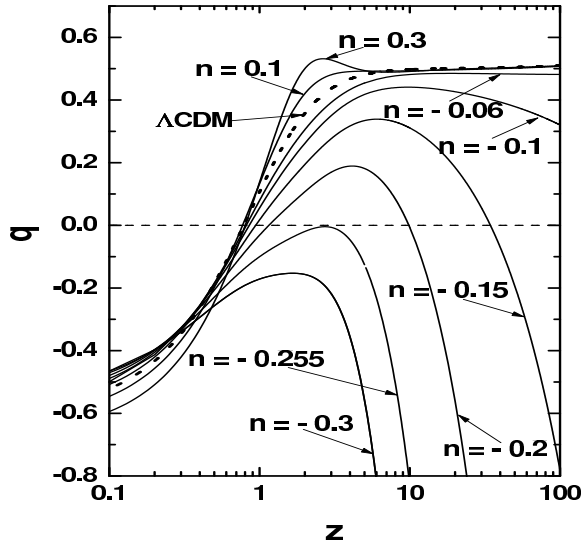,width=3.0truein,height=2.8truein}} }
\caption{Cosmic deceleration/acceleration history as a function of redshift for the $f(R)$ gravity theory given by (\ref{first-f}). Several cases are shown for selected values of $n$ and $\Omega_{mo}=0.26$.
While positive values of $n$ provides a $q(z)$ evolution very similar to the standard $\Lambda$CDM model, negative values of $n$ predict either an always accelerating universe or  two periods of acceleration separated by a deceleration phase.}
\label{q-versus-z}
\end{figure}

In Fig.~\ref{q-versus-z} we show the evolution of the deceleration parameter with $z$ for several values of $n$ and $\Omega_{mo}=0.26$, which corresponds to the best-fit value found from a joint analysis involving SNe Ia + BAO + CMB + $H(z)$ data~\cite{Janilo,Fabiocc}. As can be seen from
Fig.~\ref{q-versus-z}, for positive values of $n$, $q(z)$ approaches the standard $\Lambda$CDM evolution at intermediary redshifts and the Universe has only one recent transition toward an accelerating universe. In this case, the transition redshift does not vary considerably with $n$.

On the other hand, negative values of $n$ have great influence on the $q(z)$ evolution. In particular, we find that $f(R)$ models given
by (\ref{first-f}) with $n\leq -0.255$ predicts an always accelerating universe, which is clearly incompatible with a past dark matter-dominated epoch (and the standard $t^{2/3}$ law), whose existence is fundamental for the structure formation process to take place\footnote{This value of $n$ is slightly dependent on the matter density parameter in the [0.1, 0.3].}. Note that for $-0.255 < n <0$ these modified scenarios predict two recent periods of acceleration separated by a deceleration phase.

\begin{figure}[t]
\centerline{
{\psfig{figure=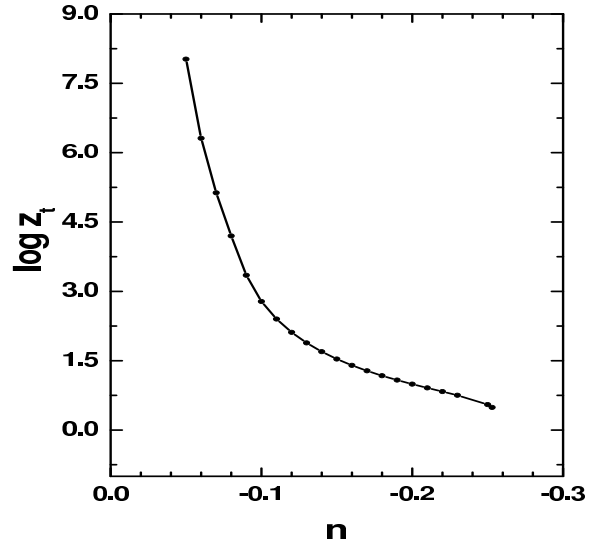,width=3.0truein,height=2.8truein}} }
\caption{The first transition redshift ($z_t$)  as a function of $n$.  Note that as $n\rightarrow 0$ the epoch of the first acceleration phase is pushed to higher $z$.}
\label{z-transiction}
\end{figure}

Fig.~\ref{z-transiction} shows the first transition redshift $z_t$ as a function of $n$. We find that the first epoch of acceleration is pushed to higher redshifts as $n$ approaches zero from negative values and that such an effect is extremely sensitive to this parameter. For instance, for $n=-0.05$ the first transition happens at very high $z$ ($z_t= 10^{8}$) whereas for $n\lesssim -0.08$ it occurs during the matter epoch ($z_t \lesssim 10^3$). It is also worth mentioning that, although we have plotted Figs.~\ref{q-versus-z} and~\ref{z-transiction} for a single value of  $\Omega_{mo}$ (= 0.26),  we have indeed repeated our analysis for a larger interval ($\Omega_{mo}\in [0.1,0.3]$). We found that the behavior of $q(z)$ and $z_t$ is very similar to the one shown in both figures, which means that for the current allowed range of the matter density parameter,  negative values of $n$ are incompatible with a past matter-dominated epoch. This means that to be consistent with cosm
 ological observations from LSS and CMB data the only correction to GR from a $f(R)$-gravity like (\ref{first-f}) is a IR modification ($n \geq 0$).  Note that an exception to this conclusion is the limit $n\rightarrow 0$ (from negative values), which for all practical purposes is equivalent to the late-time evolution of the standard $\Lambda$CDM scenario.

\section{Concluding Remarks}  \label{Conclusion}


Cosmological models based on $f(R)$ gravity may exhibit a natural acceleration mechanism without introducing a dark energy component. However, in order to provide a realistic alternative describing the late-time dynamics of the Universe, it is qualitatively well accepted (since the current observational data are inconclusive to this respect -- see Table I) that such theories must provide only an IR correction to GR. In this paper, we have discussed quantitatively this question by studying the evolution of the deceleration parameter with $z$ and imposing the existence of a past matter-dominated epoch. For the class of  $f(R)$ gravity given by (\ref{first-f}), we have shown that negative values of $n$ (high-energy correction) predict either an always accelerating universe (which is inconsistent with the past structure formation process) or  two periods of acceleration separated by a deceleration phase. These results show that even passing the background tests (such as supernova
 ) and escaping from local gravity constraints practically the entire branch of negative values of $n$ is ruled out by the cosmographic bounds discussed here.

\begin{acknowledgments}

The authors acknowledge financial support from Conselho Nacional de Desenvolvimento Cient\'{i}fico e Tecnol\'ogico (CNPq - Brazil) and Instituto Nacional de Estudos do Espa\c{c}o (INEspa\c{c}o). NP and JS also thank FAPERN for the grants under which this work was carried out.

\end{acknowledgments}

\end{document}